# Electroosmotic flow dipole: experimental observation and flow field patterning

Federico Paratore[1,2], Evgeniy Boyko[2], Govind V. Kaigala[1,*], Moran Bercovici[2,3,*]

[1]*IBM Research - Zurich, Säumerstrasse 4, 8803 Rüschlikon, Switzerland*

[2]*Faculty of Mechanical Engineering, Technion - Israel Institute of Technology, Haifa, 3200003 Israel*

*Department of Mechanical Engineering, The University of Texas at Austin, Austin, Texas 78712, USA*

*corresponding authors: M.B. (mberco@technion.ac.il) and G.V.K. (gov@zurich.ibm.com)

We experimentally demonstrate the phenomenon of electroosmotic dipole flow that occurs around a localized surface charge region under the application of an external electric field in a Hele-Shaw cell. We use localized deposition of polyelectrolytes to create well-controlled surface charge variations, and show that for a disk-shaped spot, the internal pressure distribution that arises, results in uniform flow within the spot and dipole flow around it. We further demonstrate the superposition of surface charge spots to create complex flow patterns, without the use of physical walls.

Electroosmotic flow (EOF) arises from the interaction of the ions in the electric double layer at a solid-liquid interface and an external electric field [1]. When the surface charge on the walls is inhomogeneous, non-uniform EOF arises. This phenomenon was extensively investigated in the field of capillary electrophoresis due to its undesired effect on sample dispersion [2–5]. The mechanism for dispersion was elucidated in the work of Herr *et al.* [6] which showed that as result of the non-uniform EOF, an internal pressure gradient builds up in the capillary leading to Poiseuille-type flow far from the surface charge discontinuity. Stroock *et al.* [7], using alternating stripes of different surface charges perpendicular to the electric field, showed that in the vicinity of the discontinuity mass conservation leads to flow recirculation.

The case of capillaries or microfluidic channels is a specific limit in which both the width and the depth of the fluidic configuration are significantly smaller than its length. A different limit is the case of non-uniform EOF in a microfluidic chamber where the two in-plane characteristic lengths are on the same order of magnitude. Boyko *et al.* [8] expanded on this concept and, based on previous theoretical formulations by Ajdari [9,10] and Long *et al.* [11], developed a theoretical framework to account for arbitrary zeta potential distributions on two parallel plates separated by thin-liquid film (Hele-Shaw cell). Their analysis predicted that a local non-uniformity in surface charge creates regions of higher and lower pressure, leading to in-plane recirculation, which for the case of a disk-shaped non-uniformity coincides precisely with dipole flow.

Here we present the first experimental study of non-uniform EOF in a Hele-Shaw configuration, confirming the existence of an electroosmotic flow dipole. We further show that, as expected from theory, superposition principles apply, and under combination of electric field and externally imposed uniform flow, the classical potential-flow solution of flow around a cylinder is retrieved [12,13]. This observation, in which streamlines can be curved without the use of physical walls, can potentially be leveraged as a mechanism for microscale flow patterning. In support of this, we demonstrate the ability to apply a desired surface charge distribution using polyelectrolytes, translating into the corresponding flow pattern.

Figure 1 presents a schematic of our experimental set-up designed to study the flow due to a localized surface charge non-uniformity in a Hele-Shaw chamber. The set-up consists of two parallel plates (order of $L=1$ cm) separated by a gap of $h=15$ μm containing an aqueous solution; one of the plates is patterned with $r_0=100$ μm - 500 μm diameter disk-shaped region having a modified zeta potential. In this way, the central assumption of a Hele-Shaw cell ($h \ll r_0 \ll L$), as used by Boyko *et al.*, holds. Furthermore, we use an electrolytes with ionic strengths in the range of 1 to 50 mM, for which the assumption of a thin electric double layer regime [9,10] also holds and therefore the electro-kinetic effects can be incorporated accounting for the slip velocity, which is described by the Helmholtz-Smoluchowski relation [1]

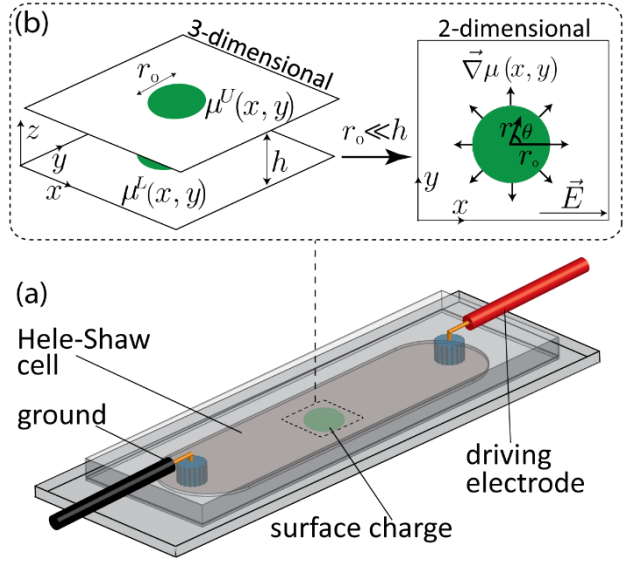

FIG. 1. *Schematic of a Hele-Shaw cell with a patterned surface charge distribution. (a) The set-up consists of two parallel plates, separated by a gap $h$. A finite region on the bottom plate is chemically altered to have a different electroosmotic wall mobility than its surrounding. An electrolyte is placed between the two plates and is subject to a uniform electric field $E$. (b) Assuming lubrication conditions (i.e. for the case $h \ll r_0$), the system can be reduce to a 2-dimensional system in which the equivalent electroosmotic distribution is the average between the upper and lower plate ones.*

$$\vec{u}^i = \mu^i \vec{E} = -\left(\varepsilon \zeta^i / \eta\right)\vec{E}. \quad (1)$$

where $\mu^i$ is the electroosmotic (EO) wall mobility, $\vec{E}$ is electric field in the $x-y$ plane and the superscript $i$ indicates the upper ($U$) and lower ($L$) plates, respectively.

Under these conditions, the depth-averaged streamline function $\psi = \psi(x,y)$ is governed by (see detailed derivation in the Supporting Information) [8]

$$\nabla^2 \psi = \left(\vec{E} \times \vec{\nabla}\bar{\mu}\right) \cdot \hat{z}, \quad (2)$$

where $\hat{z}$ is the unitary vector in the z-direction and $\bar{\mu}$ is the arithmetic mean value of the EO wall mobilities, $\bar{\mu} = (\mu^L + \mu^U)/2$. We note that $-\nabla^2 \psi$ also describes the depth-averaged vorticity.

In a polar coordinate system, $(x,y) \to (r,\theta)$, where $r(x,y)$ is the radial vector with origin at the center of the disk and $\theta(x,y)$ is the angle between $r(x,y)$ and the $x$-direction, we can solve Eq. (2) for the stream function,

$$\psi(r,\theta) = \begin{cases} -\left[\dfrac{1}{2}(\bar{\mu}_{out} - \bar{\mu}_{in})E\left(\dfrac{r_0}{r}\right)^2 - \bar{\mu}_{out}E + u_{ext}\right]r\sin\theta & r > r_0 \\ -\left[\dfrac{1}{2}(\bar{\mu}_{out} - \bar{\mu}_{in})E - \bar{\mu}_{out}E + u_{ext}\right]r\sin\theta & r \le r_0, \end{cases} \quad (3)$$

where $\bar{\mu}_{in}$ and $\bar{\mu}_{out}$ are the mean values of the EO wall mobilities inside and outside the disk, respectively. Here, $u_{ext}$ is an additional (constant) degree of freedom that can be interpreted as a uniform velocity due to other forces such as an external pressure gradient. This equation represents a family of streamlines that can be expected to arise in the presence of a localized surface charge non-uniformity (see Supplementary Figure 1).

Experimentally, one method to obtain such a surface mobility distribution is by deposition of polyelectrolytes carrying a different charge than the native surface [7]. We use FITC-labeled poly(allylamine hydrochloride) (PAH), a cationic polyelectrolytes to impose a positive charge onto the negatively charged glass (the FITC label assists to visualize the patterns). Similarly, we use poly(L-lysine) grafted with poly(ethylene glycol) side-chains (PLL-PEG) to screen the electric double layer in regions where we seek to eliminate the surface charge.

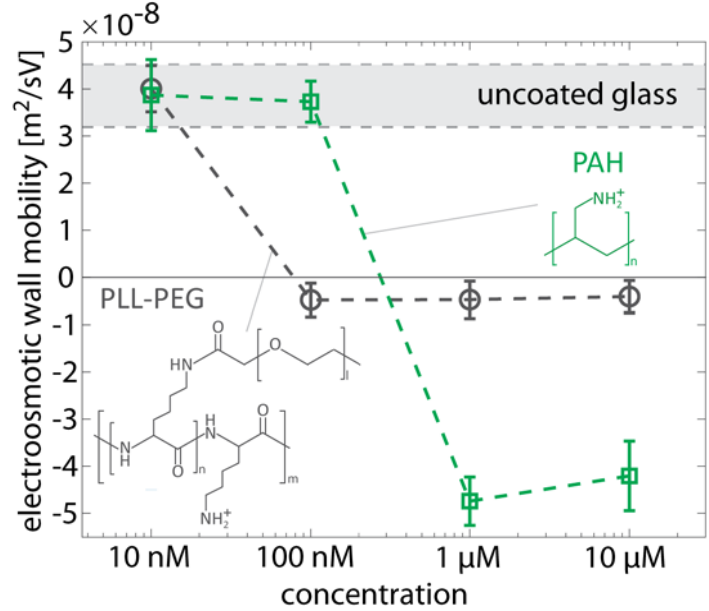

FIG. 2. *Experimental measurements of EO wall mobility of a glass/PDMS microchannel uniformly coated with PAH or PLL-PEG as a function of their concentrations in the coating solution. Both PLL-PEG and PAH exhibit a sharp transition of the EO wall mobility for concentrations higher than a threshold. For the PAH case, a concentration between 100 nM and 1 µM is required to obtain an EO wall mobility of approximately $-4.5\times10^{-8}$ $m^2/sV$, thus fully inverting the native glass EO wall mobility. For the case of PLL-PEG, a concentration in the range 10 nM to 100 nM is sufficient to reduce the EO wall mobility to approximately $4.5\times10^{-9}$ $m^2/sV$. The error bars and the grey area (uncoated channel) represent the 95% confidence interval of the mean of at least 3 different microchannels.*

Figure 2 shows the characterization of the EO wall mobility as function of the polyelectrolyte concentrations in the coating buffer (100 mM tris, 50 mM HCl, 100 mM NaCl (pH 8.2) [14] and 10 mM hepes, 5 mM NaOH (pH 7.4) [15] for PAH and PLL-PEG respectively). These experiments were performed using a 5 mM bistris, 2.5 mM HCl (pH 6.4) buffer containing neutral fluorescent 1 µm beads to trace the flow, with an applied electric field of 50 V/cm. Both polyelectrolytes show a clear transition of the EO wall mobility for concentrations higher than a threshold. Deposition of PAH with concentrations higher than a value between 100 nM and 1 µM, translates into a EO wall mobility of $-4.5\times10^{-8}$ m$^2$/sV, thus inverting the native glass EO wall mobility. Deposition of PLL-PEG with concentrations in the range 10 nM to 100 nM or higher reduces the EO wall mobility to $-4.5\times10^{-9}$ m$^2$/sV, one order of magnitude smaller than the one produced by a PAH-coated glass.

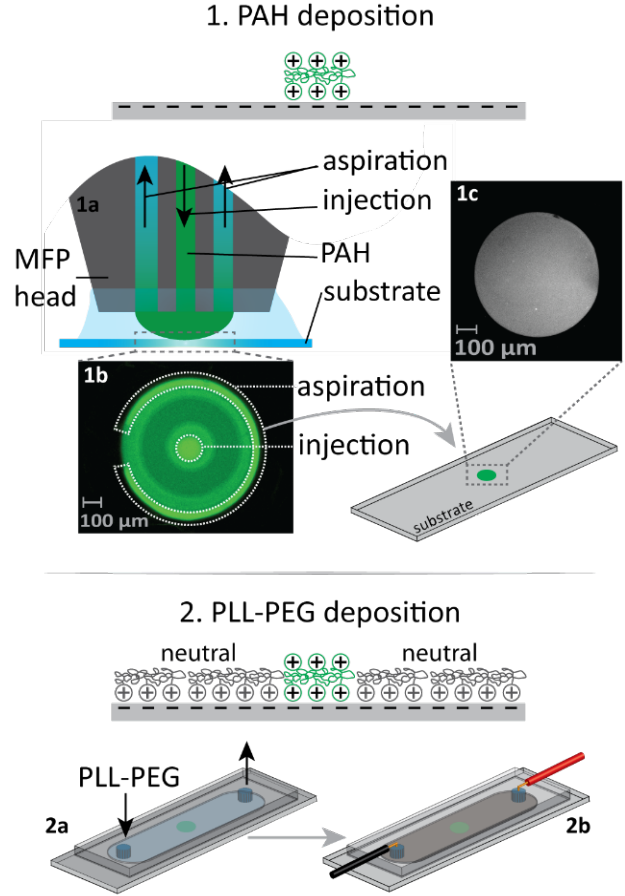

FIG. 3. *Creation of zeta potential patterns using a microfluidic probe. (1) PAH, a positively charged polyelectrolyte, is deposited locally onto a negatively charged glass slide. (1a) Working principle of a microfluidic probe. The solution containing the polyelectrolytes is injected onto the surface through an injection channel while being aspirated out at a higher flow rate by one or more adjacent aspiration channels, creating a hydrodynamic flow confinement (HFC) at the apex of the MFP head. (1b) Raw fluorescence image of axisymmetric HFC using an injection channel surrounded by an aspiration ring while depositing FITC-labeled PAH. (1c) Fluorescence image of the resulting disk-shaped PAH coated region. (2) The remaining surface is coated with PLL-PEG, a polyelectrolyte that results in an approximately zero surface charge. (2a) A microfluidic chamber is formed by a molded PDMS through which is flushed a solution containing the PLL-PEG. (2b) The chamber is filled with the working solution, and a uniform electric field is applied.*

Common methods to pattern surfaces with organic molecules (such as micromolding in capillaries (MIMIC) [16,17] and its variants [18]) use a network of channels in conformal contact with the surface, through which the coating liquid is injected, replicating the channel network geometry as a surface pattern. These approaches are suitable only for patterns with connected topologies, e.g. for channels connected to a reservoir containing the coating agent. We seek to create more complex surface patterns, including ones with unconnected topology (e.g. isolated disks). Whereas such patterns could be achieved by microcontact printing [19] or photolithography, these require dedicated microfabricated molds. Instead, we use a microfluidic probe (MFP) [20], a non-contact scanning tool that confines hydrodynamically processing liquids in a µm-sized region between the probe and the target surface. This approach enables the formation of arbitrary surface patterns using the same tool. Figure 3 and Supplementary Movie S1 show the process of patterning using the MFP.

Figure 4 presents experimental and analytical streamlines generated by the disk-shaped PAH pattern, setting a positive EO wall mobility $\bar{\mu}_{in}$ within the disk, surrounded by PLL-PEG coating, setting an essentially zero wall EO mobility $\bar{\mu}_{out} \sim 0$ outside the disk (see Supplementary Movie S2). In the absence of an externally applied flow ($u_{ext}=0$), we observe experimentally (Fig. 4(a)) the formation of an electroosmotic flow dipole outside the disk and uniform flow inside. This is in agreement with the analytical prediction (Eq. 4) where the term $-\bar{\mu}_{out}E+u_{ext}$ is set to zero (Fig. 4(b)). Our observation also confirms that the interaction between the electroosmotic flow dipole and an externally applied flow is predicted by superposition, as described by Eq. (4). Figure 4(c) shows the experimental flow field for the case of $u_{ext}=\bar{\mu}_{in}E/2$, in the direction opposing the flow within the disk; under this condition the depth-averaged velocity within the disk vanishes and, in agreement with theory (Fig. 4(d)), the streamlines curve around the spot, taking the shape of a flow around a cylinder.

Superposition also holds well for other configuration. For example, Supplementary Figure 2 shows the interaction of multiple dipoles which also can be well predicted by superposing multiple instances of solution (Eq. 4) having different origins.

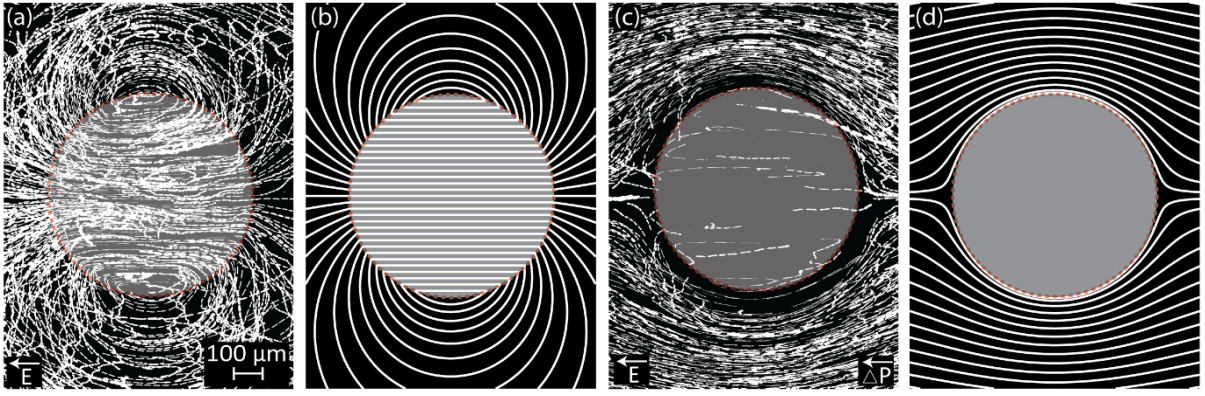

FIG. 4. *Experimental and analytical flow streamlines generated by a disk with uniform zeta potential (PAH) surrounded by a neutral surface (PLL-PEG). (a, b) With no imposed external pressure gradient, the flow is uniform in the inner region of the disk and takes the shape of a dipole in the outer region, with two vortices forming around each pole. Here the electric field is ~30 V/cm. (c, d) The addition of a pressure gradient of approximately 0.33 mbar/cm stagnates the flow within the disk, and causes the streamlines to curve around it, despite the absence of any physical walls. This flow field coincides with the solution for potential flow around a cylinder.*

These observations suggest that beyond the fundamental study of miscroscale flow due to surface non-uniformities, deliberate non-uniform charge distribution can be used to engineer desired flow fields. Figure 5(a) illustrates the concepts of 'writing' a surface charge distribution using polyelectrolyte deposition. Figure 5(b) presents the flow field resulting from a surface pattern displaying the text 'EOF!'. Surface charge discontinuities parallel to the electric field result in pressure gradients and acceleration or deceleration of the flow – this can be observed, for example, in streamlines expanding out as they move from PAH to PLL-PEG coated regions. Surface charge discontinuities perpendicular to the electric field result in creation of vorticity (as evident from Eq. (2)), observable as localized circulations over the horizontal segments of the letters. Because the pattern can be deconstructed to a set of overlapping dipoles, all effects are local, and the total mass flux of this configuration remains zero.

In this work, we investigated experimentally the non-uniform electroosmotic flow arising from surface charge non-uniformity in a Hele-Shaw cell configuration. A fundamental observation is that a spot of different surface charge gives rise to an internal pressure distribution which results in a planar electroosmotic flow dipole. We can expect that such elaborate flows would naturally occur in any surface containing charge non-uniformities under an electric field. The electroosmotic flow dipole is composed of an internal region of uniform velocity and an external region in which the velocity coincides with that of a pure dipole. Despite its difference from a pure dipole, we demonstrated that its superposition with an externally imposed flow yields the well-known solution of flow around a cylinder. Importantly this flow field, in which streamlines bend around a cylinder-shape, is obtained in an entirely unobstructed chamber.

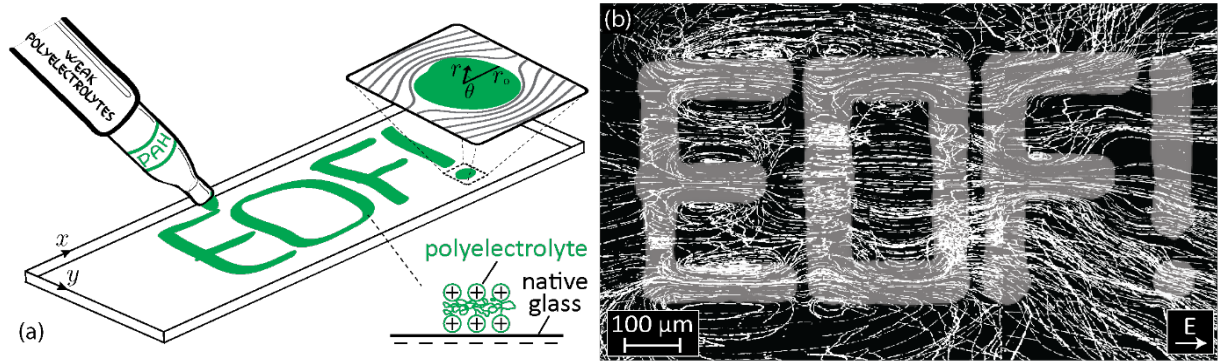

FIG. 5. *Concept of creating complex flow fields by surface charge patterning. (a) Patterning of surface zeta potential by depositing weak polyelectrolytes using a microfluidic probe (represented by a marker pen in the sketch) displaying the text 'EOF!'. (b) Visualization of the resulting flow field. The EOF on the PAH patterned regions is directed from right to left. A variety of flow patterns can be obtained using such patterning, including flow divergence and convergence, counter rotating vortices, nested vortices, shear regions, stagnation points, and inflection points. The image is constructed by stitching of 3 images, each obtained by overlaying a fluorescence streakline image of the beads (TRITC filter) with a fluorescence images of the patterned surfaces (FITC filter). Here the electric field is ~30 V/cm.*

These observations suggest that it is possible to manipulate microscale flows in a Hele-Shaw chamber solely by imposing a surface charge distribution, without the use of physical walls or mechanical components. We demonstrated this concept by using local deposition of polyelectrolytes and showed that a complex flow field can be realized. Two limitations of using polyelectrolyte deposition are that (i) only static flow configurations can be obtained and (ii) the number of zeta potential values are limited by the number of polyelectrolyte available. These limitations can likely be overcome by using gate electrodes [21] or optical means [22] to modify the surface charge dynamically, thus enabling a continuous range of surface charge values and dynamic control of the flow field. We believe that such advancements would allow the formation of tools and capabilities that simply cannot be created by conventional methods.

This work was supported by the Initial Training Network, Virtual Vials, funded by the FP7 Marie Curie Actions of the European Commission (FP7-PEOPLE-2013-ITN-607322). E.B. is supported by the Adams Fellowship Program of the Israel Academy of Sciences and Humanities. We thank X. van Kooten, D. Taylor and Dr. R. Lovchik for useful discussions. F.P. and G.V.K. acknowledge E. Delamarche and W. Riess for continuous support.

# Supplementary Material

# Electroosmotic flow dipole: experimental observation and flow field patterning

Federico Paratore[1,2], Evgeniy Boyko[2], Govind V. Kaigala[1,*], Moran Bercovici[2,3,*]

[1]*IBM Research - Zurich, Säumerstrasse 4, 8803 Rüschlikon, Switzerland*
[2]*Faculty of Mechanical Engineering, Technion - Israel Institute of Technology, Haifa, 3200003 Israel*
*Department of Mechanical Engineering, The University of Texas at Austin, Austin, Texas 78712, USA*
*corresponding authors: M.B. (mberco@technion.ac.il) and G.V.K. (gov@zurich.ibm.com)

# Table of Contents

## Supplementary Figures

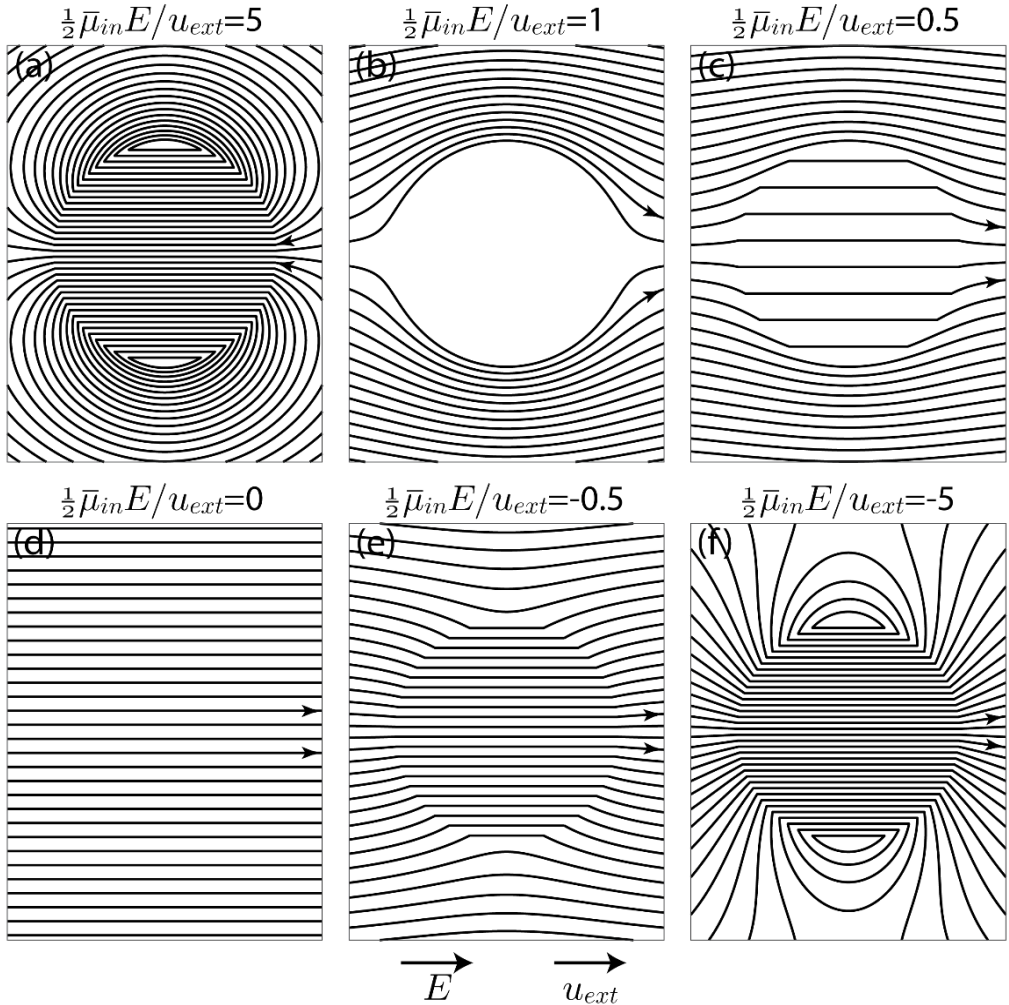

**Supplementary Figure 1.** The flow field generated by a disk-shaped patterned surface is a function of the dipole strength, $\bar{\mu}_{in}E/2$, relative to the uniform flow $u_{ext}$. When the EOF generated by the disk dominates over the uniform flow, the flow takes the shape of a dipole (**a,f**). When $\bar{\mu}_{in}E/2$ exactly equal to $u_{ext}$, no streamlines penetrate the disk-region (**b**). When the surface charge at the disk is negligible the external field dominates, resulting in uniform flow (**d**). For $0 < \bar{\mu}_{in}E/2 < u_{ext}$, the streamlines enter the patterned region bending outward (**c**), whereas, when the EOF on the disk is in the same direction of the external flow, the streamlines locally bend inwards (**e**).

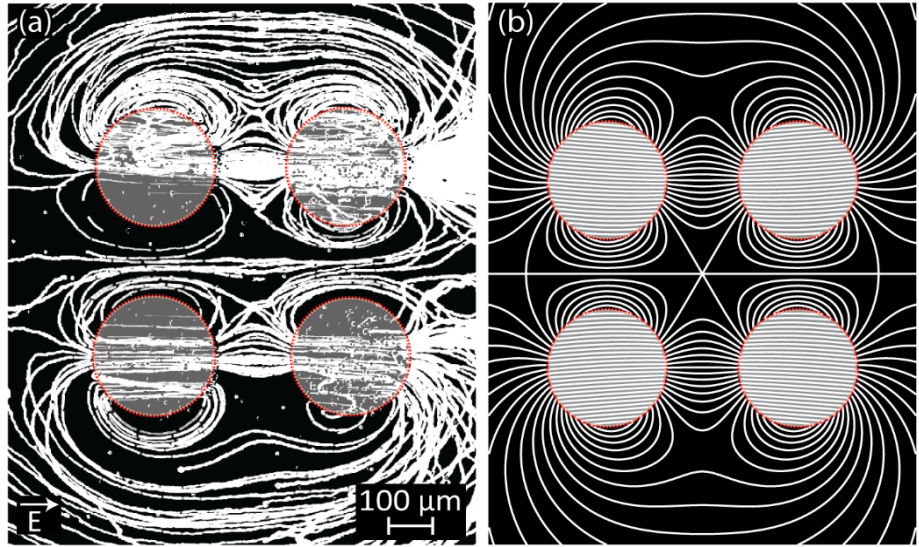

**Supplementary Figure 2.** Streamlines of a flow field generated by four disks, with centers placed 1.6 diameters apart. (**a**) Experimental measurements using 250 µm diameter PAH disks surrounded by a PLL-PEG coated surface. The image is constructed by overlaying fluorescence streakline images of randomly scattered in the flow. The red circles indicate the edges of the PAH patterned regions. Here the electric field is ~30 V/cm. (**b**) Analytical results. The measurements confirm that, owing to the linearity of governing equations, the resulting streamlines are described by the superposition of individual dipole flows.

# Supplementary Movies

**Movie S1**

Fluorescence video showing the creation of disk-shaped surface patterns using a horizontal MFP. Using an axisymmetric probe placed 20 µm above a glass substrate, we inject a solution containing PAH FITC-labeled through a central injection port at 5 µl/min while aspirating from an outer ring at 15 µl/min, thus confining the injected solution on the surface where the PAH is adsorbed onto the glass substrate. By continuously stopping the injection flow, translating the MFP and restarting the injection flow, we are able to create multiple disk-shaped surface patterns.

**Movie S2**

Fluorescent video showing the motion of neutral beads carried by a flow field generated by a disk with uniform zeta-potential surrounded by a neutral surface under an electric field. Initially, with no external pressure gradient, the beads follow the streamlines of a dipole outside the disk and are carried by a faster uniform flow in the inner region. Upon application of a pressure gradient that opposes the uniform flow within the disk, the bead trajectories bend around the disk, following the streamlines characteristic of a flow around a cylinder.

# Streamline derivation for a disk-shaped region

For completeness, here we present a summary of the theory developed by Boyko at al.[1] in dimensional form and provide a closed-form solution for the case of a disk-shaped region with uniform zeta-potential distribution. We consider the steady creeping flow an aqueous electrolyte with viscosity $\eta$ and dielectric permittivity $\varepsilon$ within a narrow gap $h$ between two parallel plates (Hele-Shaw cell configuration) with a surface zeta potential distribution $\zeta(x,y)$, subjected to a uniform in-plane electric field $\vec{E}$, in a Cartesian coordinate system $(x,y,z)$, whose $x$ and $y$ axes lie at the lower plate and $z$ is perpendicular thereto. The lower and upper plates have an arbitrary zeta-potential distribution, defined as $\zeta^L(x,y)$ and $\zeta^U(x,y)$, respectively, which vary over a characteristic length scale $l$ in the $x-y$ plane, where $h \ll l$.

In typical microfluidic configurations fluid inertia is negligible compared to viscous stresses and the transverse dimension is much smaller than the longitudinal dimensions. In addition, the thickness of the electric double layer (~1-10 nm) is much smaller than any other relevant length scales. In this low-Reynolds-number and thin-electric-double layer regime, the fluid motion in the bulk can be described by the lubrication equations,[2]

$$\vec{\nabla} p = \eta \frac{\partial^2 \vec{u}}{\partial z^2}, \quad \frac{\partial p}{\partial z} = 0, \quad \vec{\nabla} \cdot \vec{u} + \frac{\partial u_z}{\partial z} = 0, \tag{S1}$$

and the electroosmotic effects are incorporated through the use of the Helmholtz-Smoluchowski slip boundary conditions[3]

$$\vec{u}\big|_{z=0} = \mu^L \vec{E}, \quad \vec{u}\big|_{z=h} = \mu^U \vec{E}, \quad u_z\big|_{z=0,\,h} = 0, \tag{S2}$$

where we defined the electroosmotic (EO) wall mobility as

$$\mu^i = -\varepsilon \zeta^i / \eta, \tag{S3}$$

for $i = U, L$, where $U$ and $L$ refer to the upper and lower plates, respectively. Here, $p$ is the fluid pressure $\vec{\nabla} = (\partial/\partial x, \partial/\partial y)$ is the in-plane two-dimensional gradient, $\vec{u}$ is the in-plane velocity and $u_z$ is the velocity component perpendicular to it.

Integrating the in-plane momentum Eq. (S1) twice with respect to $z$ and using the slip boundary conditions Eq. (S2), we obtain the in-plane velocity field, which is then averaged vertically and leads to the depth-averaged in-plane velocity

$$\vec{u}_{aver} = -\frac{h^2}{12\eta} \vec{\nabla} p + \bar{\mu} \vec{E} \tag{S4}$$

where $\bar{\mu}$ is an arithmetic mean value of the electroosmotic wall mobility $\bar{\mu} = (\mu^L + \mu^U)/2$.

Applying the normal component of the curl operator to Eq. (S4) and using the relation $\vec{\nabla} \times \vec{\nabla} p = 0$, we eliminate the pressure and obtain an equation in terms of stream function $\psi$,

$$\nabla^2 \psi = \left( \vec{E} \times \vec{\nabla} \bar{\mu} \right) \cdot \hat{z}. \tag{S5}$$

The governing equation (Eq. (S5)) is a Poisson equation for the stream function which takes into account the effect of non-uniform electroosmotic flow through the source term depending on gradients of electroosmotic wall mobility in a direction normal to the applied electric field.

Of particular interest in the context of this work is the case of a homogeneously-coated disk-shaped surface of radius $r_0$, surrounded by a region of homogenous surface charge. Indicating with '*in*' and '*out*' the regions inside and outside the disk-shaped pattern, respectively, we define the arithmetic mean value of the electroosmotic wall mobility $\bar{\mu}_{in}$ and $\bar{\mu}_{out}$, respectively,

$$\bar{\mu}(r) = \begin{cases} \bar{\mu}_{in} & \text{inside disk} \\ \bar{\mu}_{out} & \text{outside disk} \end{cases}. \quad (S6)$$

For an electric field $E$ directed along the $\hat{x}$ axis, substituting Eq. (S6) into Eq. (S5) and using polar coordinates with the origin in the center of the disk, yields the solution for the stream function

$$\psi(r,\theta) = \begin{cases} -\left[\frac{1}{2}(\bar{\mu}_{out} - \bar{\mu}_{in})E\left(\frac{r_0}{r}\right)^2 - \bar{\mu}_{out}E + u_{ext}\right]r\sin\theta & r > r_0 \\ -\left[\frac{1}{2}(\bar{\mu}_{out} - \bar{\mu}_{in})E - \bar{\mu}_{out}E + u_{ext}\right]r\sin\theta & r \leq r_0, \end{cases} \quad (S7)$$

where $u_{ext}$ corresponds to a velocity component due to external forces acting in the $\hat{x}$ direction (e.g. external pressure gradients), $r = \sqrt{x^2 + y^2}$ is the radial coordinate and $\theta$ is the azimuthal angle.

# Experimental Procedures

## Beads coating protocol

We obtained all reagents (unless otherwise specified) from Sigma-Aldrich and prepared all solutions using DI using a Millipore system (Synergy UV, Merck Millipore). We coated 0.8-µm carboxyl fluorescent particles (Spherotech Inc.) with PLL-PEG, poly(L-lysine) randomly grafted with poly(ethylene glycol) side-chains (PLL(20)-g[3.5]-PEG(2), SuSoS AG), by the following protocol from step (i) to (v). We (i) mixed 5 mg/ml beads with 0.5 mg/ml of PLL-PEG dissolved in 10 mM Hepes/5mM NaOH in a total volume of 50 µl and (ii) let it react for 15 min at room temperature. We (ii) centrifugated them at 8000 rpm for 2 min using a standard centrifuge (5424, Eppendorf), (iv) removed the supernatant and (v) resuspended them in 200 µl of 10 mM Hepes/5mM NaOH. We repeated steps form (ii) to (v) twice. We washed the beads with DI following the protocol from step (vi) to (viii). We (vi) centrifugated them at 8000 rpm for 3 min, (vii) removed the supernatant and (viii) resuspended them in 1 ml of DI. We repeated steps form (vi) to (viii) twice, resuspended in 50 µl of DI ad stored at +4°C for maximum 10 days, after which we disposed the beads and prepared a new batch. We characterized the electrophoretic mobility of the beads by dynamic light scattering (Nano-ZS, Malvern PAnalytical), measuring a zeta potential of -4.00 ± 1.48 mV.

## MFP head fabrication

We fabricated the MFPs using standard microfabrication techniques. For the vertical MFP, we used a lithographic step to define four 50 µm-wide channels on a 4 in. silicon wafer that we etched 50 µm using DRIE. At one end of the channels, we used a second lithographic step on the same silicon wafer to define circular areas that we etched through using DRIE to form the chip-to-world fluidic connections. For the horizontal MFP, we used a lithographic step to define an aspiration port, a 600 µm internal diameter and 100 µm-wide open ring, an injection port (a 100 µm-diameter circle) located in the center of the aspiration ring, and the regions for the chip-to-world fluidic connections. We etched-through these regions using DRIE and we used a second lithographic step to define a network of microchannel to route the fluid from the aspiration and injection port to the chip-to-world fluidic connections, which we etched for 50 µm using DRIE. Finally, for both vertical and horizontal MFP, we bond the Si wafer to a glass wafer using anodic bonding, followed by dicing and polishing.

## Microfluidic chambers and channels fabrication

We fabricated PDMS (Sylgard 184, Dow Corning) microchannels and chambers using standard soft-lithography technologies. We created a mold by patterning a 15 µm layer of SU-8 on a 4 in. silicon wafer and we used a mask to define a 100 µm-wide and 2 cm-long channel (used for the EO wall mobility characterization) and a 3 mm-wide and 1.5 cm-long chamber (used for the flow patterning experiments). We fabricated the channel by pouring PDMS (using a cross-linker to monomer ratio of 1:10) on the mold to a thickness of approximately 3 mm and cured it at 60°C for 3 h.

## Surface patterning

We first cleaned a standard microscope glass slide (the substrate) by rinsing it with acetone and ethanol for 1 min each, followed by 1 min of atmospheric plasma. We dissolved Poly(fluorescein isothiocyanate allylamine hydrochloride) [Poly(allylamine hydrochloride):Fluorescein isothiocyanate 50:1] (Sigma-Aldrich), which through the manuscript we indicate as PAH, in a buffer composed of 100 mM tris, 50 mM HCl, 100 mM NaCl (pH 8.2) which maximizes its absorption onto glass[4]. We patterned the surface using a MFP with flow rates controlled by two or four syringes pumps (neMESYS 290N, Cetoni GmbH). We used a horizontal axisymmetric MFP to create disk-shaped pattern of approx. 700 µm diameter, using an aspiration rate of 15 µl/min and an injection rate of 5 µl/min. We used the vertical MFP to pattern the 'EOF!' text and the multiple dipoles. For the 'EOF!' text, we used a circular footprint of approx. 250 µm diameter achieved by setting the injection rate at 1.5 µl/min and the aspiration rate at 3 µl/min. For the multidipoles, we used a smaller circular footprint of approx. 125 µm diameter, obtained by 4 active channels forming an outer HFC using 1 µl/min injection rate and 1 µl/min aspiration rate, and an inner HFC using 1.5 µl/min injection rate and 3 µl/min aspiration rate. After patterning the surface, we rinsed the slide with DI and dried it with nitrogen gas. Separately, we cleaned a PDMS microfluidic chamber with ethanol, and then put it in contact with the patterned glass to form a closed chamber. To pattern the uncoated regions, we flushed the chamber for 5 min with PLL-PEG, in a solution of 10 mM hepes and 5 mM NaOH (pH 7.4)[5]. Finally, we rinsed the chamber with DI to remove any residual of the coating solutions.

**Experimental setup**

For the surface patterning, we used an inverted epi-fluorescent microscope (eclipse Ti−U, Nikon) equipped with a solid state light source (SOLA, Lumencor), a 10X (NA = 0.3) Nikon Plan Fluor and a 2X (NA=0.06) Nikon Plan UW objectives. We imaged using a color CMOS camera (DS-Fi3, Nikon), using an exposure time of 10 ms. We used a filter-cube (mCherry,AHF) (562/40 excitation, 641/75 nm emission, and 593 nm dichroic mirror) for imaging the beads, and a filter-cube (FITC, Nikon) (480/30 nm excitation, 535/45 nm emission, and 505 nm dichroic mirror) for imaging the PAH-fitc. For the EOF patterning experiments, we used an inverted epi-fluorescent microscope (eclipse Ti-S, Nikon) equipped with a solid state light source (Mira, Lumencor), a 10X (NA = 0.45) Nikon PlanApo λ and a 4X (NA=0.13) Nikon Plan Fluor objectives, and 0.7X de-magnification lens (Micropix Ltd.). We imaged using a CCD camera (Clara, Andor-Oxford Instrument) cooled to −40° C, using an exposure time of 1 or 3 s. We generated the electric field along the channel/chamber using a high-voltage power supply (2410, Keithley). We imposed an external pressure gradient by applying a negative pressure using a flow controeller (MFCS-EZ, Fluigent). In all the flow patterning experiments and EO wall mobility characterization, we used a 5 mM bistris, 2.5 HCl (pH 6.4) buffer and tracked the flow by using 0.8-µm fluorescent beads coated with PLL-PEG.

**EO wall mobility characterization**

To test the effect of the polyelectrolytes on the EO wall mobility, we uniformly coated a 2 cm long, 15 µm deep and 100 µm wide PDMS/glass microchannel by flushing it with PAH and PLL-PEG concentrations ranging from 1nM to 10 µM, for 5 minutes. The coating buffer is 100 mM tris, 50 mM HCl, 100 mM NaCl (pH 8.2) and 10 mM hepes, 5 mM NaOH (pH 7.4) for PAH and PLL-PEG respectively. We washed the channel using DI for 2 min and then filled with 5 mM bistris, 2.5 HCl (pH 6.4) containing beads. We applied an electric field of 50 V/cm, averaged the speed of at least ten beads which we use to calculate the EO wall mobility *via* the Helmholtz-Smoluchowski relation. We followed the same protocol for uncoated channels.

# Author Contributions

All the authors conceived the research. F.P. fabricated the devices and performed the experiments. F.P. and E.B. analyzed the experimental data and compared it to the analytical model. F.P. drafted the manuscript with inputs from all the authors. All the authors discussed the data and edited the manuscript.